\documentclass[aps,prm,reprint,superscriptaddress]{revtex4-1}
\usepackage{epsfig}
\usepackage{dcolumn}
\usepackage{epstopdf}
\usepackage{graphicx}
\usepackage{ulem}
\usepackage{color}
\usepackage[euler]{textgreek}
\begin{document}
\title{A  three-order-parameter bistable magnetoelectric multiferroic metal
}
\author{Andrea Urru}\email{\tt andrewedgard1992@live.it} 
\affiliation{Dipartimento di Fisica, Universit\`a di Cagliari, Cittadella Universitaria, Monserrato, I-09042 Cagliari, Italy}
\affiliation{Scuola Superiore Internazionale di Studi Avanzati, 
Via Bonomea 265, I-34136 Trieste, Italy}
\author{Francesco Ricci}\email{\tt frankyricci@gmail.com}
\affiliation{Institute of Condensed Matter and Nanosciences (IMCN), Universit\'e Catholique de Louvain, Chemin des \'Etoiles 8, B-1348 Louvain-la-Neuve, Belgium}
\author{Alessio Filippetti}\email{\tt alessio.filippetti@dsf.unica.it}
\affiliation{Dipartimento di Fisica, Universit\`a di Cagliari, Cittadella Universitaria, Monserrato, I-09042 Cagliari, Italy}
\affiliation{CNR-IOM, UOS Cagliari, Cittadella Universitaria, Monserrato, I-09042 Cagliari, Italy} 
\author{Jorge \'I\~niguez} \email{\tt jorge.iniguez.mail@gmail.com}\affiliation{Materials Research and Technology Department, 
Luxembourg Institute of Science and Technology, 5 avenue des Hauts-Fourneaux, L-4362 Esch/Alzette, Luxembourg}
\affiliation{Department of
Physics and Materials Science, University of Luxembourg, 41 Rue du
Brill, L-4408 Belvaux, Luxembourg}
\author{Vincenzo Fiorentini}\email{\tt vincenzo.fiorentini@gmail.com}
\affiliation{Dipartimento di Fisica, Universit\`a di Cagliari, Cittadella Universitaria, Monserrato, I-09042 Cagliari, Italy}
\begin{abstract}
Using first-principles calculations we predict that the layered-perovskite metal Bi$_5$Mn$_5$O$_{17}$ is  a  ferromagnet,  ferroelectric, and ferrotoroid which may realize the long sought-after goal of a room-temperature ferromagnetic single-phase multiferroic  with large, strongly coupled, primary-order polarization and magnetization. 
Bi$_5$Mn$_5$O$_{17}$ has two nearly energy-degenerate ground states with mutually orthogonal vector order parameters (polarization, magnetization, ferrotoroidicity), which can be  rotated globally by switching between ground states. Giant cross-coupling magnetoelectric and magnetotoroidic effects, as well as optical non-reciprocity, are thus expected. Importantly, Bi$_5$Mn$_5$O$_{17}$  should be thermodynamically stable  in O-rich growth conditions, and hence experimentally accessible. 
\end{abstract} 
\date{\today}
\maketitle

\noindent{\bf Introduction}

 \noindent One of the key goals   of  multiferroics research \cite{multif}, not yet achieved after several decades, is finding a room-temperature single-phase multiferroic with large polarization and
magnetization  primary orders--i.e. neither being a weak side effect of other  phenomena.
In this context, a multiferroic metal would be of the utmost interest as the seat of robust 
magnetism,   enabling stable and large magnetization and polarization at 
application-relevant temperatures, and hence a possible path to the above goal. Also, such a material 
would quite likely be a ferromagnet, whereas most insulating magnets are antiferromagnetic.

Further,  if  additional orders \cite{schmid}
such as e.g. ferrotoroidicity were to exist (as they do in appropriate symmetry), the mutual couplings of the various orders   (e.g. magnetoelectricity) may  be quite out of the ordinary. On the other hand, more than two concurrent orders rarely coexist in a multiferroic, and this is especially true of metals, where multiferroicity itself is already unexpected (ferromagnetism in metals is common, but ferroelectricity is exceedingly rare \cite{bito}).   
 
It is therefore against all expectations that in this paper  we predict  a specific instance of a  multiferroic metal  as a possible path to the key goal of multiferroicity: the orthorhombic layered perovskite Bi$_5$Mn$_5$O$_{17}$ (BiMO henceforth) is a   multi-order-parameter, bistable, magnetoelectric, metallic, room-temperature multiferroic. Indeed, BiMO is a metal possessing three space-or\-tho\-go\-nal vector order pa\-ra\-me\-ters: magnetization {\bf M}, polarization {\bf P}, and  ferro\-to\-roi\-dal moment {\bf T}, generated by simultaneous time reversal and inversion symmetry  breaking;  it  exists in two nearly energy-degenerate multiferroic ground states, which can be transformed into one another, causing the order-parameter triad to rotate in space; it exhibits  giant  magnetoelectricity, and potentially other couplings among the three orders, including toroidicity-related optical effects; finally, it has  a sizable thermodynamic stability window, so it can be grown in practice. \\

\noindent{\bf Results}\\

\begin{figure}[ht]
\centering
\includegraphics[clip,width=1\linewidth]{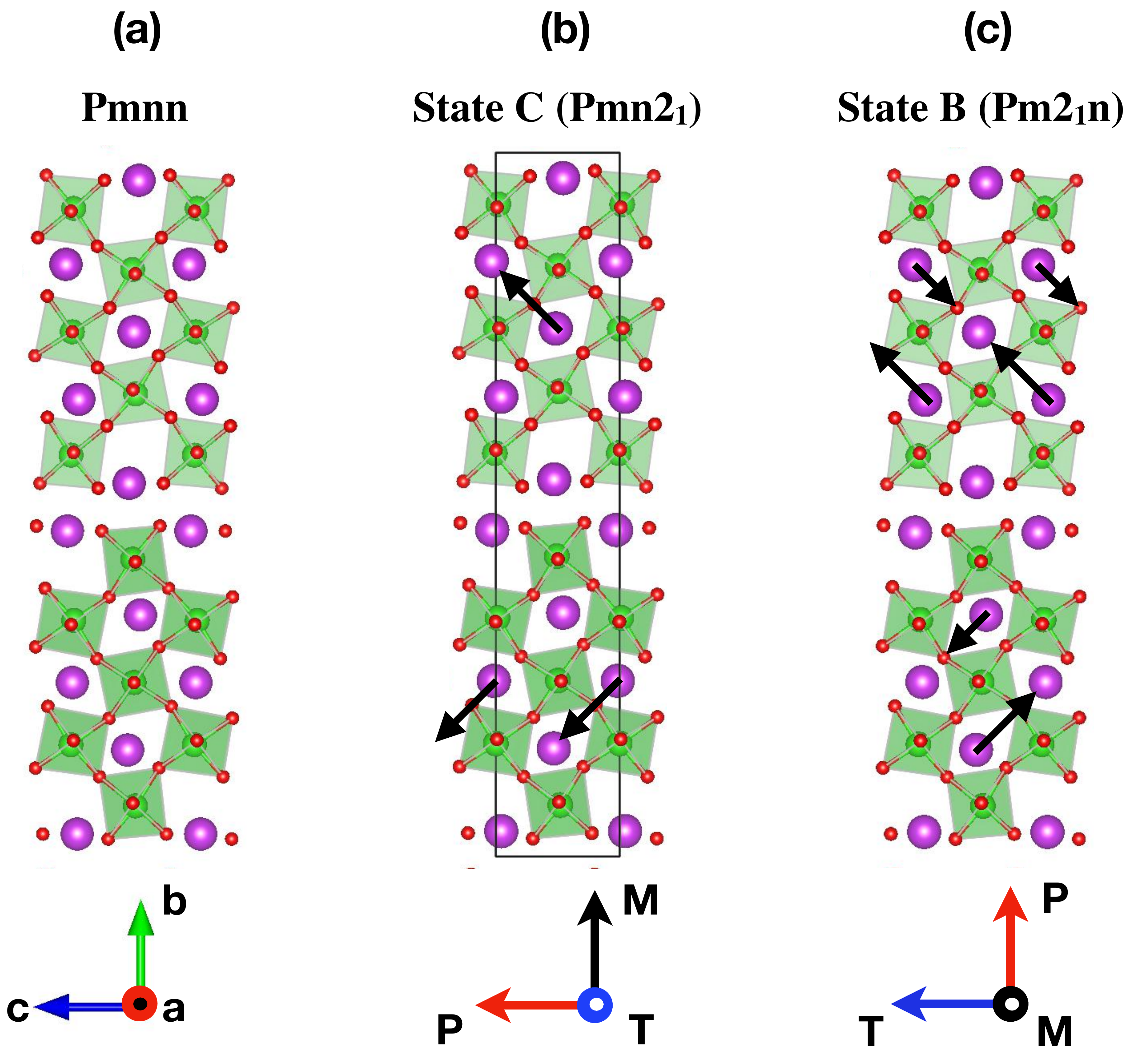}
\caption{{\bf Comparison of high-symmetry and distorted structures.}
 High-symmetry unstable (\textbf{a}) and  distorted ground-state stable (\textbf{b}), (\textbf{c}) structures of BiMO. The crystal axes, the main local ionic dipoles, and the order-parameter vectors in the different states are shown.}
\label{figstruct}
\end{figure}

\noindent{\bf Structure. } BiMO is a layered perovs\-kite of the class A$_n$X$_n$O$_{3n+2}$ with $n$=5. Its structure is depicted in Fig. \ref{figstruct}. The periodic cell compri\-ses two 5-perovskite-unit blocks along the ${\bf b}$ axis (the crystal axes are  {\bf a}=[100], {\bf b}=[011], {\bf c}=[0$\bar{1}$1] in the cubic perovskite setting).

We search for instabilities  in the {\it q}=0 phonon spectrum of the  centrosymmetric structure with $Pmnn$ space group, {computed both via density functional perturbation theory \cite{dfpt_review}  and by finite differences, with completely consistent results.} The two dominant unstable modes, with one-dimensional irreps $B_{1\text{u}}$ and $B_{3\text{u}}$,  condense into polar stable ground states with space groups $Pmn2_1$ and $Pm2_1n$.

In both phases (Fig.  \ref{figstruct}, see also Supplementary Discussion for structural data) the  Bi atoms move within the {\bf bc}-plane, and what distinguishes the structures is the modulation of the Bi displacements from layer to layer.
{A schematic representation, with indicative arrows (not to scale) corresponding to the largest dipoles, is given in Fig.  \ref{figstruct}.}
In the $Pmn2_1$ phase (C state in the following), the displacements along {\bf b} form an anti-polar pattern; instead, the displacements along {\bf c} are in phase, originating a {\bf c}-polarized distortion. 
{In the $Pm2_1n$ state, labeled B  in the following, an anti-polar pattern appears along {\bf c}; along {\bf b}, an uncompensated anti-polar pattern originates a  {\bf b}-polarized distortion.}
 All  displacements are invertible and allow for hysteresis; indeed, as discussed below, BiMO supports a depolarizing field. A third unstable mode (irrep $B_{2\text{u}}$)  leading to a structure with  symmetry $P2_1mn$ is preempted by the C and B modes, due to its much lesser energy gain.

\begin{figure}[ht]
\centering
\includegraphics[clip,width=1\linewidth]{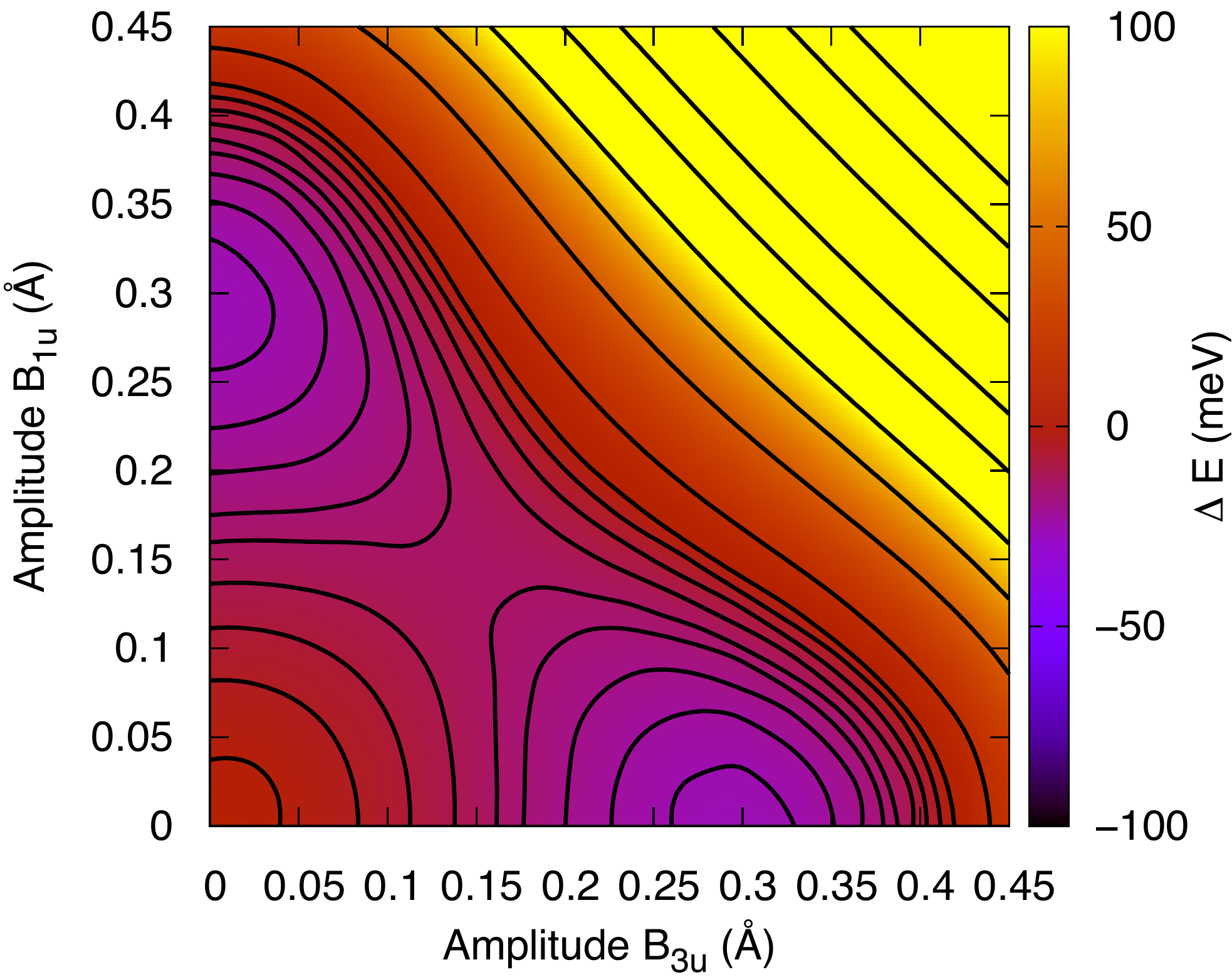}
\caption{{\bf Energy landscape for BiMO vs ferroelectric distortions.} Energy of BiMO as function of the distortion along the two main unstable (and both polar) modes of the high-symmetry phase.}
\label{figland}
\end{figure}

{The calculated energy gain upon condensing into the
  C or B state is the same, 100 \textmu eV \AA$^{-3}$, to within 1\%. The
  energy landscape in Fig.  \ref{figland} shows pictorially the two
  distinct minima and provides an estimate of about 50 \textmu eV \AA$^{-3}$
  for the barrier between the two states.  Since this energy barrier
  is similar to those occurring in other polar perovskites, we can
  estimate the ferroelectric Curie temperature at well above
  ambient. More precisely, in prototypical ferroelectric perovskite
  BaTiO$_{3}$, equivalent rhombohedral minima are separated by an
  orthorhombic saddle point, the energy barrier being about
  15 \textmu eV \AA$^{-3}$ \cite{kingsmith94}, which results in an
  orthorhombic-rhombohedral transition temperature of 183~K. Since
  such a transition temperature is known to scale with the mentioned
  energy barrier \cite{wojdel,abrahams68}, and the polar distortion in
  layered perovskite BiMO is ultimately not very different from those
  occurring in its perovskite counterparts, we can estimate a Curie
  temperature for BiMO exceeding 500 K.
BiMO will thus be locked in either ground  state C or B  at  room temperature, and can be thermally activated between them with a modest $T$ increase.}

\noindent{\bf Polarization.} Given their polar symmetry, both the B and C states of BiMO can
 possess a spontaneous polarization {\bf P}. The coexistence of metallicity and polarization has been discussed at length in our previous work on the ferroelectric metal Bi$_5$Ti$_5$O$_{17}$ \cite{bito},  a layered perovskite to which BiMO bears close similarities. As in that case,  we calculate {\bf P} with a modified Berry phase technique \cite{bito} which exploits the flatness of the bands along the polar axes and the sheet-like Fermi  surface (see below the discussion of the band structure). 
  In the  B state, {\bf P}$_{\rm B}$$\|${\bf b} and $|${\bf P}$_{\rm B}|$=0.71 \textmu C cm$^{-2}$; in the C state,  {\bf P}$_{\rm C}$$\|${\bf c} and   $|${\bf P}$_{\rm C}|$=5.03 \textmu C cm$^{-2}$. Both values are in the  same league  as III-V nitrides and II-VI oxides  (e.g. 2.9 \textmu C cm$^{-2}$ for GaN \cite{XN}).  The  electronic polarization contributions by   valence majority, valence minority, and conduction electrons  are roughly in the ratio 30:10:1.

\begin{figure}[ht]
\centering
\includegraphics[clip,width=1\columnwidth]{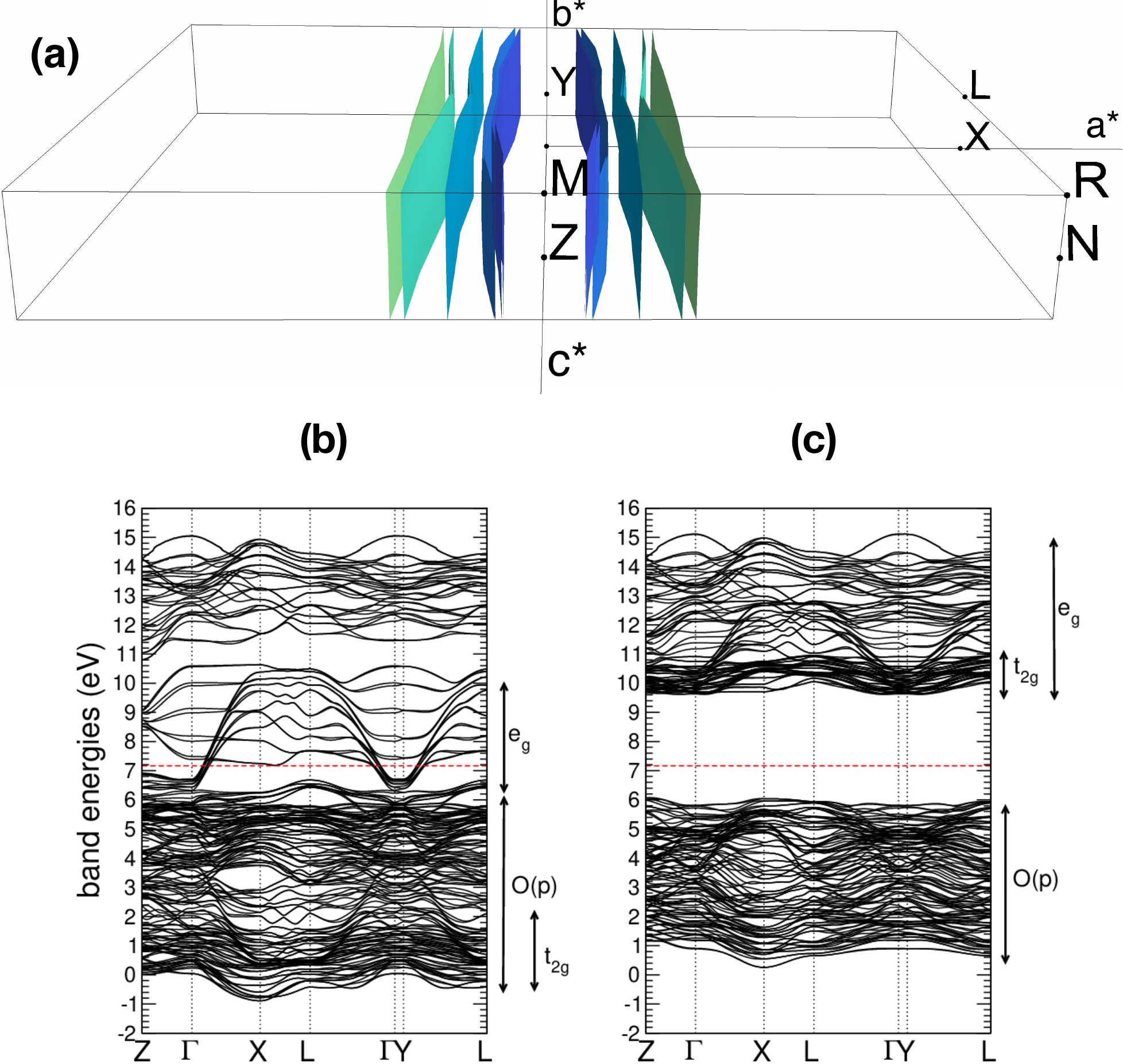}
\caption{{\bf Main features of the electronic properties of BiMO.} (\textbf{a}) 3D view of the BiMO Fermi surface; (\textbf{b}) majority and (\textbf{c}) minority bands of BiMO, calculated with the VPSIC method. Fermi level: horizontal dashed line.}
\label{bande}
\end{figure}

\noindent{\bf Bands and magnetism. } 
BiMO  is a metal, whose Fermi surface (Fig.  \ref{bande}) shows  line-like sections along the {\bf b}$^*$ and (less cleanly) {\bf c}$^*$ reciprocal axes (i.e. in   {\bf b}$^*${\bf c}$^*$-like 
 planes in the Brillouin zone), justifying the applicability of the approach of Ref.\cite{bito} to computing the polarization of both states B and C. The bands (computed including self-interaction corrections, which drastically improve predicted gaps \cite{vpsic,vpsic1}) show that  BiMO is  a half-metallic ferromagnet with a large minority gap (GGA results are the same except for the smaller  minority gap). This is expected from its nominally 3$d^{3.2}$ Mn ions coupled via  double exchange \cite{dopmn}.

The average magnetization is 3.06 \textmu $_{\text{B}}$ per Mn (2.84 \textmu $_{\text{B}}$ per Mn within atomic spheres) 
in GGA, and 3.4 \textmu$_{\text{B}}$ per Mn from VPSIC  self-consistent occupations. The energy difference of the ferromagnet and (approximate)  G-type antiferromagnet provides an  average 
{ magnetic coupling $J$$\simeq$16 meV; 
applying a Hubbard U correction with the typical Mn value U=3 eV, $J$ becomes about 24 meV (see Supplementary Discussion).}

\begin{figure}[ht]
\centering
\includegraphics[clip,width=1\linewidth]{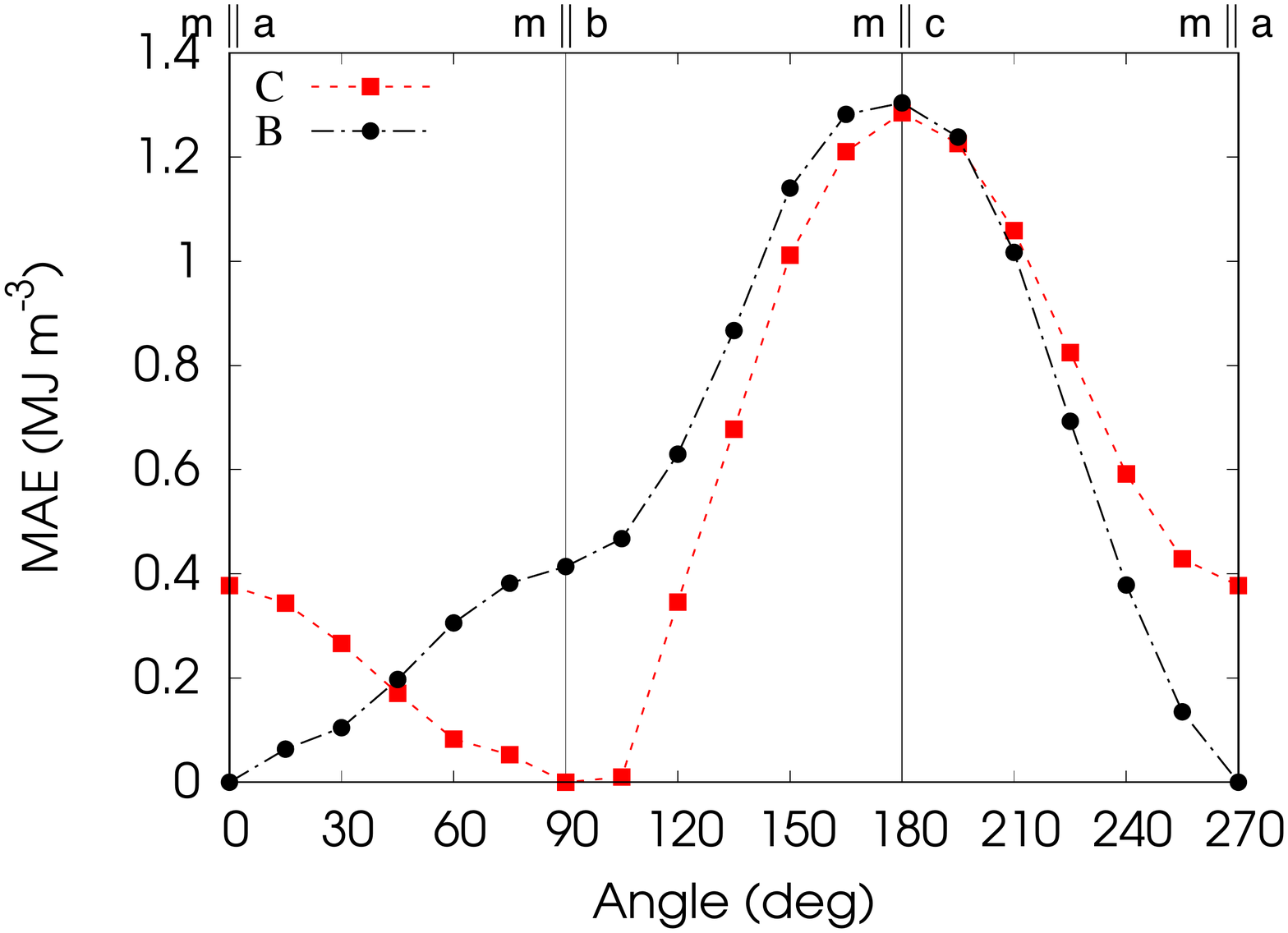}
\caption{{\bf Magnetic anisotropy in BiMO.} Energy as function of the orientation of the magnetization   for the C  (dashed, squares) and B (dash-dot, circles) ground states of BiMO. The easy axis is {\bf b} for the C state, and {\bf a} for the B state.}
\label{anisotropia}
\end{figure}

We performed non-collinear spin-orbit calculations to ascertain the orientation of the 
magnetization.  As shown in Fig.  \ref{anisotropia}, in the C state {\bf M} is parallel to the {\bf b} axis, with significant magnetoanisotropy energy barriers of
0.38 MJ m$^{-3}$ for the {\bf a}  axis  and 1.28 MJ m$^{-3}$ 
 for the  {\bf c} axis. 
For the B state,   {\bf M} is instead parallel to the {\bf a} axis (Fig. \ref{anisotropia}), with magnetoanisotropy barriers 0.41 MJ m$^{-3}$  for the {\bf b}  axis  and 1.30 MJ m$^{-3}$
 for the  {\bf c} axis. In both ground states, therefore,  {\bf M} is orthogonal to the polar axis and hence to {\bf P}  (the polarization was discussed above).

{ We now provide  estimates of the magnetic Curie temperature $T_{\text{C}}^{\bf M}$ 
based on literature numerical simulations of the Ising as well as classical and quantum  Heisenberg models \cite{heisen,heisen1}; the magnetoanisotropy energy just calculated is neither especially small or large \cite{anisotrop}, so it is not obvious which model  is preferrable.   The values of   $T_{\text{C}}^{\bf M}$ for the $J$ mentioned earlier (16 meV for U=0 eV and 24 meV for U=3 eV) are as follows: classical Heisenberg 361 K and 521 K, quantum Heisenberg 263 K and 387 K, Ising 378 K and 555 K.
  It seems  reasonable to conclude that BiMO's  $T_{\text{C}}^{\bf M}$ is near or above room temperature.}

\noindent{\bf Toroidicity. }
The fact that {\bf M} is orthogonal to {\bf P} in  both the C and B ground states agrees 
\cite{bilbao,bilbao1,schmid} with their  respective magnetic point groups being  
 $m'm2'$  and $m2'm'$;  symmetry further implies \cite{bilbao,bilbao1} that there exists a non-zero ferrotoroidal  moment 
 {\bf T} \cite{es}, the order parameter of a ferrotoroidal state \cite{schmid,ferroicT,sax,hlinka,hlinka2,spaldin-me}.   {\bf T}  
 is proportional to $\sum_i$${\bf r}_i$$\times$${\bf m}_i$, and is akin to an angular momentum with  magnetic moments functioning as  velocities; in a toroid, {\bf T} is the sum of individual current-loop terms with  {\bf r}=0 at the center of the torus, hence the name. 
 
  Symmetry also implies that the three order-parameter vectors  {\bf M}, {\bf P}, and {\bf T}  must be mutually orthogonal. Similarly to polarization, the toroidal moment is defined as a difference between two states \cite{es} (our reference structure has point group $mmm$ and hence zero moment); to obtain a well-defined {\bf T}, one must remove  toroidicity quanta  analogous to polarization quanta   \cite{es}, and  the cell must be  recentered to the average of the magnetic moments positions. Once that is done, we find   for the C state  {\bf T}$\|${\bf a} and 
{$|${\bf T}$|$$\simeq $0.27 \textmu$_{\text{B}}$\AA}, 
 and for the B state, {\bf T}$\|${\bf c} and $|${\bf T}$|$$\simeq$0.77 \textmu$_{\text{B}}$\AA, smaller than e.g. the 1.75 \textmu$_{\text{B}}$\AA\, for LiCoPO$_4$ \cite{es}, but certainly not insignificant.
 This establishes {\bf T} as the third order parameter of BiMO,  and confirms  it to be orthogonal to {\bf P} and {\bf M} in both states C and B, as dictated by symmetry (a similar symmetry-determined situation occurs in other layered perovskites such as V-doped La$_2$Ti$_2$O$_7$ \cite{notaT2} the magnetic group is $2$, and indeed  {\bf M}$\|${\bf P}$\|${\bf T}).

We note in passing that  {\bf T} will be non-zero as long as the symmetry is polar, and time reversal  is broken; thus {\bf T} would exist  even if {\bf P} were  suppressed by electronic screening (which it isn't, as discussed below); also, the symmetry of BiMO forbids the existence of the fourth `electromagnetic' order parameter, the ferroaxial or electrotoroidal moment \cite{hlinka,hlinka2}.
 
\noindent{\bf Magnetoelectricity and other  consequences. }
{ Based on the above results, BiMO should exhibit a
  number of unique properties and effects.  First and foremost, it may
  undergo multiple magnetoelectric switching, with one order parameter
  potentially switching under the field conjugate to another order
  parameter. This can be realized by a
  trilinear coupling term {\bf T}$\cdot$({\bf P}$\times${\bf M}) in
  the Landau free-energy expansion (see the Supplementary
  Discussion): it is both allowed by symmetry
  \cite{sannikov,sannikov1} and consistent with our three orthogonal vector
  order parameters. (The symmetry of the $Pmnn$ reference structure bars instead {\bf P}{\bf M}-like
  bilinear terms.)  The trilinear coupling implies  that  ground states B and C should exist in four
  distinct degenerate states 1$\equiv$(+,+,+), 2$\equiv$(--,--,+),
  3$\equiv$(--,+,--), 4$\equiv$(+,--,--), where the signs characterize
  the order parameters in our fixed set of crystallographic axes, and, for
  example, state (--,--,+) has order parameters {\bf --P}, {\bf --M},
  {\bf T}. We indeed verified directly that these states do exist,
  with calculations analogous to those in Fig. \ref{anisotropia}. It follows that, for example, switching
  {\bf P} by an electric field in state 1 will lead either to state 2
  (magnetization co-switching) or 3 (toroidicity co-switching); and
  {\bf M} switching by a magnetic field leads state 1 to either 2 or
  4. Interestingly, as briefly discussed in the Supplementary Discussion,
  our calculations suggest that {\bf T} is a secondary (slave) order
  that follows the primary orders {\bf P} and {\bf M} according to {\bf
    T}$\sim${\bf P}$\times${\bf M}. Hence, of the mentioned
  switching possibilities, the ones we expect will occur are
  1$\rightarrow$3 and 1$\rightarrow$4.}

{Another class of switching possibilities involves transitions
  between the two ground states B and C, which entail
  a space rotation of the vector-order-parameter triad, such as {\bf
    T}$\|{\bf a}$, {\bf M}$\|{\bf b}$, {\bf P}$\|{\bf
    c}$$\,\,\Rightarrow\,${\bf M}$\|{\bf a}$, {\bf P}$\|{\bf b}$, {\bf
    T}$\|{\bf c}$ for the C to B transformation. This transition could
  be driven in several ways: one could electrically pole {\bf P} from
  {\bf c} to {\bf b}, which should rotate {\bf M} from {\bf b} to {\bf
    a}, and {\bf T} from {\bf a} to {\bf c}; or more interestingly, a
  magnetic field coercing {\bf M} from {\bf b} to {\bf a} could turn
  {\bf P} from {\bf c} to {\bf b}, and {\bf T} from {\bf a} to {\bf c}
  as well.
 Since there are four degenerate orientational states in both C and B, there are  16 possible C-to-B transitions: for example, a {\bf b}$\rightarrow${\bf a} magnetization rotation could turn state C (+,+,+) into  B (+,+,+), but, due to  the degeneracy just discussed, it could also land it into, say, B (--,+,--). It is likely that such transitions will be set apart by different energy barriers, which are however extremely difficult to estimate.}

Going further, BiMO should exhibit linear static magnetoelectricity, such as $\delta${\bf M} = $\widetilde{\alpha}$ {\bf E}, measured by the magnetoelectric tensor $\widetilde{\alpha}$. According to symmetry, only its off-diagonal elements are non-zero \cite{schmid}, and specifically $\alpha_{bc}$, $\alpha_{cb}$ for state C and $\alpha_{ab}$, $\alpha_{ba}$ for state B (this is similar to the weak-ferromagnet La$_2$Mn$_2$O$_7$ \cite{pippo}, whose non-zero tensor elements are $\alpha_{bc}$, $\alpha_{cb}$). 
  Thus, for example, BiMO in state C will exhibit magnetoelectric cross-coupling  $\delta M_b$=$\alpha_{bc}$$E_c$, so that a {\bf c}-oriented electric field  changes the magnetization along {\bf b} (conversely, a {\bf b}-oriented magnetic field would cause a {\bf c}-polarized response). 
   Transforming BiMO to state B, the non-zero elements will be different and so will the cross-coupling, namely $\delta M_a$=$\alpha_{ab}$$E_b$, etc. (an important practical consequence of bistability). 
  
    Interestingly, due to the trilinear coupling, magnetoelectric coefficients have an antisymmetric component proportional to {\bf T}  in addition to the usual symmetric components \cite{es,spaldin-me,ferroicT}.  The off-diagonal response also turns out to be related to vanishing of both the ferroaxial moment \cite{bilbao,bilbao1,schmid,spaldin-me} and the magnetoelectric monopole (which is easily verified to be zero) \cite{monopoliz}.

  Another expected effect in BiMO is  optical non-reciprocity, also known as optical-diode effect \cite{optlorentz,optdiode} (see also \cite{ferrotor} for a review); this is basically tunable and switchable birefringence, visible in magneto-optical absorption \cite{optdiode} and second-harmonic generation (which was used in \cite{ferroicT} to establish the ferroic nature of the toroidal order, including hysteretic behavior). It requires non-zero toroidal moment and off-diagonal magnetoelectricity \cite{optdiode2}, both of which BiMO possesses. Such effects, expected e.g. for beams propagating along opposite directions in a toroidic material,  may also occur in BiMO under inversion of   {\bf T}, which can be effected via {\bf M} inversion under a magnetic field, e.g. from state 1 to state 4 of a given ground state as described earlier.  Additionally, in BiMO a transformation  between ground states (C and B) would enable switchable multidirectional  birefringence. In passing, we note that the more exotic linear toroidoelectric and toroidomagnetic effects \cite{schmid} are also possible in this symmetry.

We finally roughly estimate  the linear magnetoelectric coupling taking  state C as an example. The  {\bf P} and {\bf M}  changes (with respect to the centrosymmetric phase)
  $\Delta$$M$=0.24 \textmu$_{\text{B}}$ per cell and $\Delta$$P$=4.5 \textmu C cm$^{-2}$ provide a
 rough estimate of the linear coupling $\alpha$=$\Delta P$/$\Delta M$=12 \textmu s m$^{-1}$, which is  large compared to values in boracites \cite{rivera} or phosphates \cite{rivera2}. One   also obtains
$\partial M$/$\partial E$$\sim$$\Delta M$/$\Delta E$=${\chi_{\text{d}}}{\alpha}$ 
and $\partial P$/$\partial H$$\sim$$\Delta P$/$\Delta H$=$\chi_{\text{m}} \alpha,$
equal to, respectively, 3$\times$10$^{-5}$ and 2$\times$10$^{-8}$ in SI units, assuming  the  dielectric susceptibility $\chi_{\text{d}}$$\simeq$40 \textepsilon$_0$  (see below) and a characteristically "large" magnetic susceptibility  $\chi_{\text{m}}$=1000 \textmu$_0$.

\noindent{\bf Interface monopoles and  depolarizing field. }   
The discontinuity of BiMO's  zero-field {\bf P} at the interface with an unpolarized insulating medium 
should produce \cite{xnint} a sheet-like charge and hence a  depolarizing field. If this were the case,  BiMO's {\bf P}  would be switchable by an external field (see Ref.\cite{bito}). This may be  preempted, however,  by  conduction charge  compensation,  or by the disappearance of the polar distortion if the depolarizing field were too strong. Thus, to elucidate the possibility of switching BiMO's polarization, we study an insulator-cladded  BiMO layer  in a gated-device configuration.

\begin{figure}[ht]
\centering
\includegraphics[clip,width=1\linewidth]{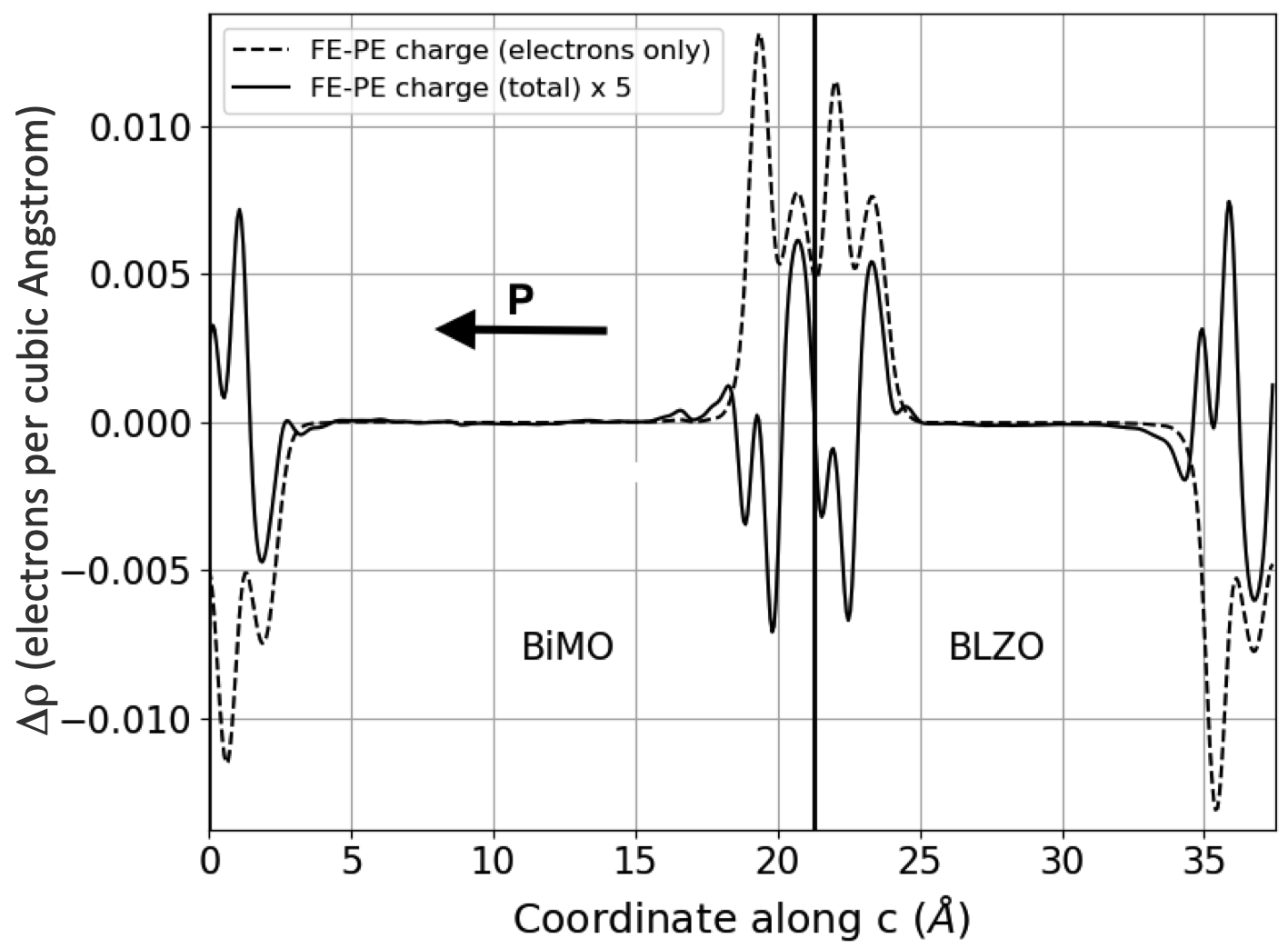}
\caption{{\bf Interface charges in BiMO/insulator superlattice.} Filtered averages of  electronic charge and of total charge in the BiMO-BLZO SL, showing polarization-originating charge accumulation at the interfaces.  Solid line and  $y$ axis mark the geometric interfaces. Negative charge drawn as positive.}
\label{monopole}
\end{figure}

\begin{figure}[ht]
\centering
\includegraphics[clip,width=1\linewidth]{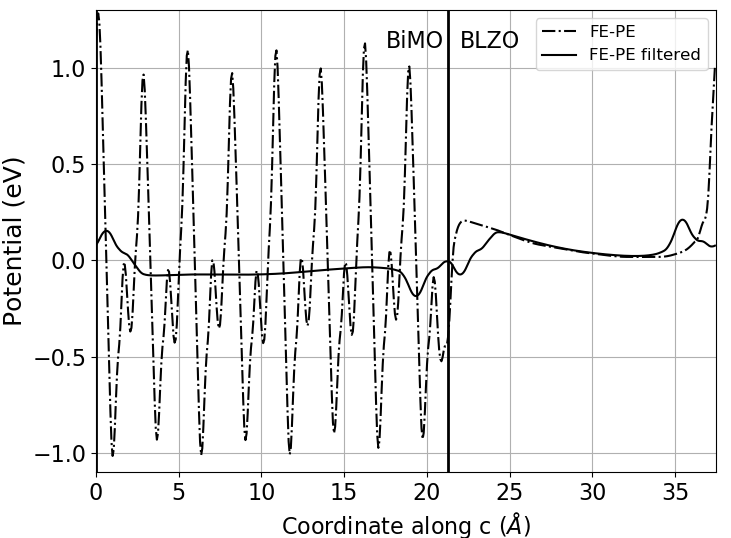}
\caption{{\bf Electric fields in BiMO/insulator superlattice.} Filtered average of potential difference of polar-distorted  and non-polar SLs, highlighting the sawtooth shape of the potential and the existence of a depolarizing field.}
\label{field}
\end{figure}

Previously \cite{bito} we showed that Bi$_5$Ti$_5$O$_{17}$ supports a field when stacked with an insulator along its polar axis {\bf b}; given the similarly flat Fermi surface of BiMO, we expect that state B will shadow that behavior closely, and we do not address it explicitly. We instead turn to   state C, which is the harder case  since, as  {\bf P}$\|${\bf c}, we need  a metal/insulator superlattice (SL) along {\bf c}, which requires a layered-perovskite $n$=5 insulator (other claddings, including vacuum, would be highly prone to interface states). 
 For our proof-of-concept
simulations, we adopt the  fictitious $Pmnn$ compound   BaLa$_4$Zr$_5$O$_{17}$  (BLZO), whose gap is 2 eV in GGA.

In Figs. \ref{monopole}  and \ref{field} we report   charge and po\-ten\-tial dif\-fe\-ren\-ces  between the non-polar  $Pnnm$ BiMO/$Pnnm$ BLZO  and  polar $Pmn2_1$  BiMO/$Pnnm$ BLZO SLs (3/4 BiMO/BLZO cells, symmetric interfaces, 378 atoms).   The macroscopically-averaged \cite{bito,filter} charge difference  shows a monopole, matching  the polarization direction, that integrates to  0.19 \textmu C cm$^{-2}$ (about 1/25 the bare {\bf P}). This  confirms that BiMO carries a  non-zero {\bf P} which the conduction charge is unable to  screen out \cite{bito} 
(the  conduction density of  3$\times$10$^{21}$ cm$^{-3}$ is low, but  still   20 times the needed screening density). The BiMO layer has a finite effective dielectric constant of $\varepsilon_{\rm BiMO}$$\simeq$40 (Eq.4, Ref. \cite{xnint} with $\varepsilon$=$\varepsilon_{\infty}$=5 for lattice-frozen BLZO), a suppression of the static Drude divergence being  admissible from  general features of metal-insulator SLs \cite{park}.   BLZO is non-polar and interfaces are symmetric, so the interface charge   must stem from  BiMO's polarization.

Accordingly, the polar--nonpolar SL potential difference  (Fig. \ref{field})  
has a slope, i.e. an electric field inside both BiMO and the cladding, whose  only possible source is the polarization within the BiMO layer. The field is 190 MV m$^{-1}$ in the insulator; 
the field of 200 MV m$^{-1}$ from the interface monopole is in the same ballpark. 
Aside from its precise value, the depolarizing field confirms that the non-zero {\bf P} of BiMO in state C survives   as  in the BiTO  B-like state (which, we recall, BiMO also possesses in addition to the C-like state being discussed). Hence, BiMO   qualifies as a multiferroic  metal. 

The screened field (energy density 1 \textmu eV \AA$^{-3}$) cannot remove the polar distortion  (100 \textmu eV \AA$^{-3}$ energy density gain), whereas the unscreened field (energy density 700 \textmu eV \AA$^{-3}$) would. Thus, similarly to BiTO \cite{bito}, we may label BiMO a self-screening hyperferroelectric magnetic metal, since  polarization survives in the thin film  thanks to self-screening (although of course the underlying mechanism in  BiMO is quite unlike that in hyperferroelectrics proper \cite{hf}).

\begin{figure}[ht]
\centering
\includegraphics[clip,width=1\linewidth]{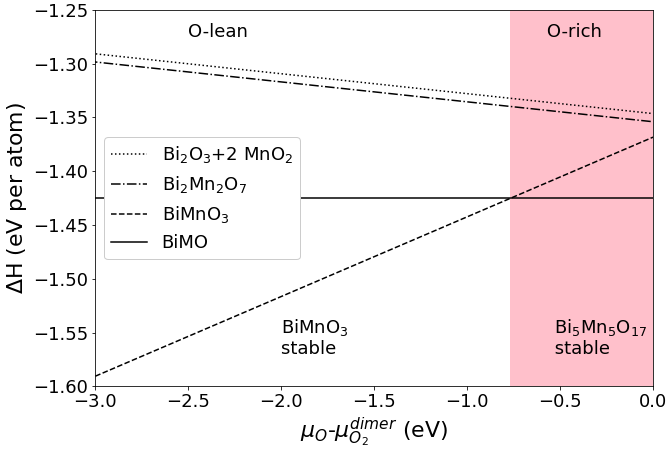}
\caption{{\bf Thermodynamic stability of BiMO.} Formation enthalpies (from O$_2$, Bi and Mn metals) of BiMO, BiMnO$_3$, Bi$_2$Mn$_2$O$_7$, and  Bi$_2$O$_3$+2$\,$MnO$_2$ vs O chemical potential.  BiMO stability region: pink area on the right.}
\label{chempot}
\end{figure}

\noindent{\bf Thermodynamic stability. }
To assess BiMO's stability within equilibrium thermodynamics, in Fig.  \ref{chempot} we compare its enthalpy of formation (see Methods) with that of a few possible alternative Bi-Mn-O systems, specifically BiMnO$_3$ (a rare insulating ferromagnet, paraelectric in equilibrium, ferroelectric under  strain \cite{dieg}), Bi$_2$Mn$_2$O$_7$ (a layered-perovskite ferroelectric and antiferromagnetic insulator, not synthesized so far), and a combination of the two binaries Bi$_2$O$_3$ and MnO$_2$ in their  most stable versions vs the chemical potential of oxygen. Mn$_2$BiO$_5$, also considered, is not competitive. 
These Bi-Mn-O  combinations are both oxygen-rich and oxygen deficient (with 3, 3.5, and 3.5 oxygen atoms per perovskite  stoichiometric  unit) compared BiMO (3.4 oxygens per unit). Clearly BiMO is the most stable of this group in an appreciable range of oxygen-rich conditions. While the stoichiometries considered here are not exhaustive, there is good circumstantial evidence for the possible stability of BiMO.

In summary, we have predicted that  Bi$_5$Mn$_5$O$_{17}$  is a multiferroic  metal
 featuring three space-orthogonal vector ferroic order parameters, and with two degenerate ground states where the order-parameter triad gets rotated in space. As such, the material   is expected to exhibit multistate multiferroicity, non-reciprocity effects, and giant magnetoelectricity. Importantly, it has a thermody\-na\-mi\-cal stability window that should make it  experimentally accessible.
This material could be the
realization of a long sought-after goal, a room-temperature
single-phase multiferroic with large and strongly-coupled
 polarization and magnetization. \\

\noindent{\bf Methods}\\

\noindent{\bf Computational details. } First-principles density-functional calculations  in the generalized gradient (GGA) and local density (LDA) approximations to density-functional theory are performed with VASP \cite{vasp,vasp1,vasp2,vasp3} {and Quantum Espresso (QE) \cite{qe,qe1}}, and supplemented with variational pseudo-self-interaction-corrected (VPSIC) calculations \cite{vpsic,vpsic1}. Structural instabilities and magnetic properties are studied in  GGA {and LDA}, and VPSIC is used for improved  electronic structure and polarization properties.  In the VPSIC code we use scalar-relativistic 
ultrasoft pseudopotentials  \cite{uspp} with plane-wave cutoff of 476 eV; in VASP we 
use scalar-relativistic projector augmented waves  \cite{paw,paw1} (valence electrons: Bi 5$d$,6$s$,6$p$; Mn 3$d$,4$s$; O 2$s$,2$p$; PAW 
data sets {\tt Bi}, {\tt Mn},  {\tt O\_s}) and a cutoff of 500 eV; in QE we use fully relativistic  ultrasoft potentials with cutoff 90 Ry. Brillouin zone integration is done on a 6$\times$2$\times$4 grid for self-consistency and optimization, and a 12$\times$4$\times$8 grid to compute  densities of states.  The 
electronic polarization  is computed with the Berry phase approach \cite{berry} as modified 
in Ref.\cite{bito}, on an 8$\times$4 set 
of 11-point k-strings along the polarization axes (i.e. a reoriented 8$\times$4$\times$11 grid). Non-collinear magnetism calculations, including spin-orbit effects,  have been double checked with VASP and QE. Convergence tests have been performed for each code separately, and structural relaxations were redone independently with both. More information in the Supplementary Methods. Further details on the potential-filtering and polarization-calculation procedures are in Ref.\cite{bito}. Formation enthalpy  data (except the layered perovskites, which we computed directly) are from the Materials Project, {\tt https://materialsproject.org}, with material IDs  
{\tt mp-23262} (Bi$_2$O$_3$),  {\tt mp-19395} (MnO$_2$), {\tt mp-35} (Mn), {\tt mp-23152} (Bi);
 {\tt mp-504697} (Mn$_2$BiO$_5$), {\tt mp-23477} (BiMnO$_3$).\\

\noindent{\bf Data availability}\\

\noindent Input files and details of procedures can be provided upon request.\\

\noindent{\bf Code availability}\\

\noindent The main codes used are public domain (QE), licensed (VASP), or custom in-house (VPSIC). The latter can be provided in the frame of a scientific collaboration. Further minor postprocessing codes can be provided upon request.\\

\onecolumngrid
\newpage
\noindent{\bf Acknowledgments} 

\noindent Work  supported in part by UniCA, FdS, RAS via Progetti di ateneo 2016 {\it Multiphysics approach to 
thermoelectricity} (VF) and 2020 {\it Stability and defects of hybrid perovskites} (AF, VF); CINECA, Italy, via ISCRA grants (VF);   Luxembourg National Research Fund, Grant No. INTER/ANR/16/11562984 EXPAND (J\'I);
Project MIUR-PRIN 2017 TOPSPIN (AF).
VF is currently on secondment leave at the Embassy of Italy, Berlin, Germany; views expressed herein are his own and  not necessarily shared by the Italian Ministry of Foreign Affairs.\\

\noindent{\bf Author contributions}

\noindent AU, VF, and JI: structure and its instabilies. AF, AU, FR, and VF: electronic, 
magnetic, and transport properties. AF: polarization properties. AU and VF: 
magnetic non-collinear properties. FR and VF: behaviour of finite layers and 
slabs. JI and VF: thermodynamic stability. AU, JI, and VF: Landau free energy 
and toroidicity. VF: main draft of the paper. AU, AF, FR, JI, VF: refinement and 
finalization of the paper. \\

\noindent {\bf Additional information}\\

\noindent{\bf Supplementary Information} accompanies this paper at http://www.nature.com/naturecommunications\\

\noindent{\bf Competing interests:} The authors declare no competing  interests.\\

\noindent{\bf Reprints and permission information} is available online at http://npg.nature.com/reprintsandpermissions/\\

\noindent{\bf How to cite this article}: Urru, A., Ricci, F., Filippetti, A. et al. A three-order-parameter bistable magnetoelectric multiferroic metal. Nat Commun 11, 4922 (2020). https://doi.org/10.1038/s41467-020-18664-6

\end{document}